\documentclass[graybox]{svmult}
\usepackage[left=1.7cm, right=1.7cm, top=2cm, bottom=2cm]{geometry}
\usepackage{caption}
\usepackage{type1cm}        
\usepackage{makeidx}         
\usepackage{graphicx}        
\usepackage{multicol}        
\usepackage[bottom]{footmisc}
\usepackage{natbib}
\usepackage{booktabs}
\usepackage{newtxtext}       %
\usepackage[varvw]{newtxmath}   
\usepackage{subfigure}

\makeindex             

\begin{document}

\title*{Inestability presented in the estimating of the Nelson-Siegel-Svensson model}
 \titlerunning{Inestability in the Nelson-Siegel-Svensson model} 
\author{Ainara Rodr\'iguez-S\'anchez}
\authorrunning{A. Rodr\'iguez} 
\institute{A. Rodr\'iguez \at Department of Applied Economics and Economic History, National University of Distance Education, Madrid, Spain. \email{arsanchez@cee.uned.es}}
%
%
\maketitle

\abstract{The literature shows the possible existence of a problem called collinearity in both Nelson-Siegel and Nelson-Siegel-Svensson models due to the relationship between the slope and curvature components. The presence of this problem and the estimation of both models by Ordinary Least Squares would lead to coefficients estimates that may be unstable among other consequences. However, these estimates are used to make monetary policy decisions. For this reason, it is important to try mitigating this collinearity problem. Consequently, some authors propose traditional procedures for the treatment of collinearity such as: non-linear optimisation, to fix the shape parameter or ridge regression. Nevertheless, all these processes have their disadvantages. Alternatively, a new method with good properties called raise regression is proposed in this paper. Finally, the methodologies are illustrated with an empirical comparison on Euribor Overnight Index Swap and Euribor Interest Rates Swap data between 2011 and 2021.\\}

\section{Introduction}
The term structure is usually measured using default-free, annualized zero-coupon yields. However, the term structure is not directly observable from the published coupon bond prices and yields. Therefore, it is necessary to use a model estimating this term structure with high accuracy since it will be used in monetary policy, among other purposes.

The term structure estimation could be done with McCulloch's cubic spline \citep{mcculloch1971measuring}, but this method has the drawback that the estimated rates can be quite unstable, especially in the longer terms \citep{shea1984pitfalls, svensson1994estimating}. Other methods exist to estimate the yield curve such as the stochastic models of \cite{vasicek1977equilibrium} and  \cite{cox1985intertemporal}, the latter being an extension of the former. Nevertheless, it is necessary to assume that volatility is constant in these models.  As a result, these are usually used only to estimate short-term rates. 

The need for a parsimonious model to estimate the term structure curve was recognized by \cite{friedman1977time} when he stated "Students of statistical demand functions might find it more productive to examine how the whole term structure of yields can be described more compactly by a few parameters". As a consequence, a parsimonious linear model was suggested by \cite{nelson1987parsimonious} and later by \cite{svensson1994estimating}. 

The Nelson-Siegel model and its extension, the Nelson-Siegel-Svensson model, are widely used by central banks and others market participants to estimate the term structure. However, \cite{james2000interest,choudhry2010fixed} suggested that there is a limitation in this first model, as it is not flexible. The Nelson-Siegel model is accurate only for yield curves with a hump, not for those with a hump and a trough.

On the other hand, certain authors \citep{annaert2013estimating,annaert2015estimating,diebold2006forecasting,leon2018multicollinearity} suggested that Nelson-Segel and Nelson-Siegel-Svesson models could presented a collinearity problem due to there may be a relationship between the explanatory variables in both linear models. If these models with a collinearity problem are estimated using Ordinary Least Squares (OLS), the estimates obtained may be unstable with signs contrary to expectations and difficulties may be encountered in interpreting the significance values, among other important consequences \citep{marquaridt1970generalized,willan1978meaningful,Gunst1977b,Farrar1967}.

For the treatment of the collinearity problem, the literature proposed the application of nonlinear optimization methods. However,  \cite{virmani2012estimability, leon2018multicollinearity} showed that the models are very sensitive to the initial value used in optimization. In addition, some authors turned the static Nelson-Siegel model into a dynamic model. As a result, \cite{barrett1995yield,fabozzi2005predictability,diebold2006forecasting} proposed to fix the shape parameter over the whole time series of the term structures. Nevertheless, this is not insignificant as it can result in an extremely smooth time series of the parameter estimates \citep{annaert2013estimating}. 

Finally, \cite{leon2018multicollinearity,annaert2013estimating,annaert2015estimating} applied ridge regression, an alternative methodology to OLS, traditionally used to overcome the collinearity problem and proposed by \cite{hoerl1970ridgea, hoerl1970ridgeb}. Although, ridge regression has some disadvantages such as the decrease in the coefficient of determination and the impossibility of using global and individual significance test \citep{Rodriguez2019}. Alternatively, this paper proposes the application of raise regression which maintains the original characteristics of the model (in relation to the goodness of fit and the global significance of the model) while mitigating the collinearity. In \cite{nelsoncapitulo} is used this methodology to estimate the Nelson-Siegel model but focus on mitigating the collinearity problem. Whereas in this manuscript, the main goal is to minimize the Mean Square Error (MSE) by trying to mitigate the collinearity problem.

The structure of the paper is as follows: Section 2 presents a theoretical first look at the  Nelson-Siegel and Nelson-Siegel-Svensson models. Section 3 introduces the collinearity problem, including its detection tools, and the Mean Square Error. Section 4 reviews the different methodologies traditionally applied to treat multicollinearity in the Nelson-Siegel and Nelson-Siegel-Svensson models. Section 5 the main contribution of this paper is showed: the application of the raise regression. In Section 6, the proposal contribution is applied for one empirical example compared to OLS and ridge regression. Finally, Section 6 summarizes the main conclusions.

\section{An overview of the Nelson-Siegel and Nelson-Siegel-Svensson models}\label{Nelsonteoria}

In this section, the model is introduced starting from the Nelson-Siegel model. Following  \cite{annaert2013estimating}, the spot rate function $r(\tau, \lambda)$ at time to maturity $\tau$ is specified as:

\begin{equation}\label{nelson}
r(\tau, \lambda) = 
\left[\begin{matrix}
\beta_{0} \ \beta_{1} \ \beta_{2}\\
\end{matrix}\right]
\left[\begin{matrix}
1\\
\frac{\lambda(1-e^{-\frac{\tau}{\lambda}})}{\tau}\\
\frac{\lambda(1-e^{-\frac{\tau}{\lambda}})}{\tau} - e^{-\frac{\tau}{\lambda}}
\end{matrix}\right]
=
\beta_{0} r_{0} + \beta_{1} r_{1} + \beta_{2} r_{2}\\
\end{equation}

where $r_{0}$, $r_{1}$ and $r_{2}$ represent the level, slope and curvature components of spot rate curve, respectively. Also, the shape parameter, $\lambda$, determines the location of the hump and the parameter $\beta_{2}$ the relevance of curvature. On the other hand, the economic interpretation is as follows: the long-term component is  $\beta_{0}$ and the short-term component is $\beta_{0}+\beta_{1}$.

The Nelson-Siegel model describes term structures with a single curvature. However, market data may have a second curvature, generally at a longer maturity. In order to solve this problem, \cite{annaert2015estimating} suggested using the Nelson-Siegel-Svensson model as follow:

\begin{equation}\label{siegel}
r(\tau,  \lambda) = 
\left[\begin{matrix}
\beta_{0} \ \beta_{1} \ \beta_{2}\ \beta_{3}\\
\end{matrix}\right]
\left[\begin{matrix}
1\\
\frac{\lambda_{1}(1-e^{-\frac{\tau}{\lambda_{1}}})}{\tau}\\
\frac{\lambda_{1}(1-e^{-\frac{\tau}{\lambda_{1}}})}{\tau} - e^{-\frac{\tau}{\lambda_{1}}}\\
\frac{\lambda_{2}(1-e^{-\frac{\tau}{\lambda_{2}}})}{\tau} - e^{-\frac{\tau}{\lambda_{2}}}
\end{matrix}\right]
=
\beta_{0} r_{0}+ \beta_{1} r_{1} + \beta_{2} r_{2} + \beta_{3} r_{3}\\
\end{equation}

where the shape parameters, $\lambda_{1}$ and $\lambda_{2}$, determines the location of the first and second hump, respectively. Moreover, the parameter $\beta_{3}$ represents the relevance of the second curvature. 

Thus, these models present a structure similar to a general linear regression with $n$ observations and $(p-1)$ independent variables as follows:
\begin{equation}\label{regresion}
\mathbf{Y}=\mathbf{X}\boldsymbol{\beta}+\mathbf{u},
\end{equation}
where \textbf{u} is a random disturbance (that is assumed to be spherical with variance $\sigma^{2}$), $\mathbf{X}_{nx(p-1)}$ is the matrix of observations of the independent variables ($\mathbf{X}_{0} = (1,1,...,1)^{t}$), and $\mathbf{Y}_{nx1}$ is the vector of the observations of the dependent variable. Finally, the dynamic form of model \ref{regresion}, whose estimated parameters may vary over time, is considered in this paper \citep{diebold2006forecasting}. As a result, model \ref{nelson} and model \ref{siegel} are estimated $m$ times.

\section{The nature of the multicollinearity problem and mean square error}\label{multicolinealidadnelson}

\cite{gilli2010calibrating,annaert2013estimating,annaert2015estimating,diebold2006forecasting,leon2018multicollinearity} showed that there may be a collinearity problem caused by the relationship among the slope and curvature components of the Nelson-Siegel and Nelson-Siegel-Svensson models. When this problem exists, it is not recommendable to estimate the linear model with OLS since it has undesirable consequences such as inflated variances of estimated parameters. 

As mentioned before, both models are used for monetary policy and other important issues. For this reason, it is relevant to detect and treat this collinearity problem. Then, the traditional Condition Number (CN) proposed by \cite{besley1980regression} is used to diagnose the possible collinearity problem as follows:
\begin{equation}
    \label{NC.def}
    K(\mathbf{X}) = \sqrt{\frac{\xi_{max}}{\xi_{min}}},
\end{equation}
where $\xi_{max}$ and $\xi_{min}$ are the maximum and minimum eigenvalues of matrix $\mathbf{X}^{t} \mathbf{X}$, respectively. Note that before calculating the eigenvalues, the data of the $\mathbf{X}$ matrix are transformed into unit length. The aim of this transformation is to obtain a column-equilibration, as this results in a more meaningful CN. Moreover, unit length scaling is similar to standardized scaling, used to transform the $\mathbf{X}^{t} \mathbf{X}$ matrix into a correlation matrix, which leads some to confuse the two issues \citep{belsley1991guide,salmeron2018transformation}. However, the standardized scaling is inappropriate for analyzing the collinearity because it tends to produce misleading diagnoses of conditioning.  When the standardized scaling is used the data are centered, obtaining a zero mean, so the constant term disappears. \cite{salmeron2019noesencial} and \cite{salmeron2018variance} show that the CN with standardized data does not detect non-essential collinearity (linear relationship among the constant term and some independent variable of the linear model) but only detects essential collinearity (linear relationship two or more independent variables of the linear model, excluding the constant term). Finally, \cite{besley1980regression} and \cite{salmeron2018variance} establish that a value of CN  between 20 and 30 indicate moderate collinearity while values higher than 30 implies a strong collinearity. 

An additional tool to detect non-essential collinearity is the coefficient of variance (CV) of the independent variables. CV values below 0.1002 suggest that the variable has minimal variability, indicating a strong association with the constant term \citep{salmeron2019noesencial}. Conversely, the traditional Variance Inflation Factor (VIF) is used to detect essential collinearity problems when it exceeds 10 \citep{Salmeron2018Garcia}.

Another no less important problem is to have a high bias between the observed value and the estimated value. Therefore, the main objective of this contribution is to reduce the Mean Square Error (MSE) while trying to mitigate collinearity problem in the background. This MSE, derived from the Sum Squared Errors of each estimated model ($SSE_{j}$), is expressed as:
\begin{equation}
    \label{ECM.def}
    MSE= \frac{1}{m} \sum \limits ^{m}_{j=1} SSE_{j},
\end{equation}
where $SSE_{j} =\sum \limits ^{n}_{i=1}  \left (r(j, i)-\hat{r} (j, i) \right)^{2}$, $r(j, i)$ is the observed value and $\hat{r} (j, i)$ is the estimated value.

\section{Traditional estimation procedures}

Since the importance of both minimizing the MSE and mitigating the collinearity problem in the Nelson-Siegel and Nelson-Siegel-Svensson models has already been mentioned in the previous section, this section will focus on the traditional estimation processes used to mitigate collinearity and its disadvantages:

\begin{itemize}
\item  \cite{virmani2012estimability,cairns2001stability} proposed the application of nonlinear optimization methods but these are very sensitive to the initial value used in optimization.
\item  \cite{barrett1995yield,fabozzi2005predictability,diebold2006forecasting} suggested to fix the shape parameter over the whole time series of the term structures in the estimation of the Nelson-Siegel model. The first and second authors fix this parameter to 3 for annualized returns, and the last fixing the shape parameter at 0.0609 for monthly returns. However, it can result in an extremely smooth time series of the parameter estimates.
\item  \cite{leon2018multicollinearity,annaert2013estimating,annaert2015estimating} used ridge regression methodology which consists in mitigating the collinearity problem from a numerical point of view by introducing a positive quantity denoted as $k$ in the main diagonal of matrix $\mathbf{X}^{t}\mathbf{X}$ \citep{hoerl1970ridgea, hoerl1970ridgeb}. Moreover, note that the condition number of the ridge regression ($K(\mathbf{X}, k)$) should be calculated by following \cite{salmeron2018transformation}. Nevertheless, this methodology has certain disadvantages, such as the requirement for using standardized data \citep{garcia2016standardization, salmeron2017note, Rodriguez2019}. If standardized data are used in the estimation of the Nelson-Siegel-Svensson model would result in the loss of an important economic parameter, the long-term ($\beta_{0}$). 
\end{itemize}

\section{Proposed estimation process}\label{raiseregression}

The method proposed called raise regression was introduced by \cite{garcia2011raise} and further developed by \cite{salmeron2017raise,gomez2021raise}. In this case, this methodology consists in mitigating the collinearity problem from a geometric point of view. Given that collinearity arises when the vectors are very close geometrically, the most effective approach for treating this problem involves augmenting the angle among both vectors. 

In a model such as the one given in the expression (\ref{regresion}), the residuals of the auxiliary regression $\mathbf{X}_{i} = \mathbf{X}_{-i}\boldsymbol{\delta} + \mathbf{v}$ (where $\mathbf{X}_{-i}$ is the matrix $\mathbf{X}$ after eliminating the independent variable $\mathbf{X}_{i}$), denoted as $\mathbf{e}_{i}$ are used to raise variable $i$, with $i=1,\dots,(p-1)$, as $\widetilde{\mathbf{X}}_{i} = \mathbf{X}_{i} + k \mathbf{e}_{i}$ with $k \geq 0$ (raising factor), and it is verified that $\mathbf{e}_{i}^{t} \mathbf{X}_{-i} = \mathbf{0}$, where $\mathbf{0}$ is a vector of zeros with appropriate dimensions. In this case, the raise regression consists of the OLS estimation of the following model:
    \begin{equation}
        \mathbf{Y} = \widetilde{\mathbf{X}} \boldsymbol{\beta}(k) + \widetilde{\mathbf{u}},
        \label{modelo_alzado}
    \end{equation}
     where $\widetilde{\mathbf{X}} = [\mathbf{X}_{0}  \ \mathbf{X}_{1} \dots \widetilde{\mathbf{X}}_{i} \dots \mathbf{X}_{(p-1)}] = [\mathbf{X}_{-i} \  \widetilde{\mathbf{X}}_{i}]$. \cite{garcia2011raise} showed that this technique does not alter the overall characteristics of the initial model. In \cite{gomez2021raise}, various criteria for selecting which independent variable to raise are presented. This manuscript will used two of these criteria: raising the variable with the highest VIF (it limit tends to 1) and  raising the variable with the lowest CV (it will increase).

Note that the condition number of the raise regression ($K(\widetilde{\mathbf{X}}, k)$) should be calculated by following \cite{de2020analysis}. Also, to calculate the Sum Squared Errors of the raise regression ($SSE(\widetilde{\mathbf{X}}, k)_{j} $) should follow \cite{gomez2021raise}. The SSE of the raise regression coincides with the SSE of the OLS. Therefore, MSE of the raise regression will be the same as MSE of the OLS.

On the other hand, this paper proposes an alternative methodology for estimating the  Nelson-Siegel and Nelson-Siegel-Svensson models focused on minimizing the MSE and on mitigating collinearity in the background. The process and steps to follow for estimation are similar to those used for the estimation of ridge regression in \cite{annaert2013estimating, annaert2015estimating}:

\begin{enumerate}
\item Perform a grid search based on the OLS regression to obtain the estimate of $\lambda$ or ($\lambda_{1}$, $\lambda_{2}$) which generates the lowest $SSE_{j}$.
\item Calculate the condition number for the optimal $\lambda$ or ($\lambda_{1}$, $\lambda_{2}$).
\item Re-estimate the coefficients by using raise regression only when the condition number is above a specific threshold, in this case 20. The raise constant is chosen using an interactive searching procedure that finds the lowest positive number, $k$, which makes the recomputed condition number falls below the threshold. As a result, the correlation between the regressors will decrease and so will the condition number.
\end{enumerate}

  \section{Empirical comparison of the estimation methods} \label{example}
  
To illustrate the estimation procedures of the previous sections in both Nelson-Siegel (NS) and Nelson-Siegel-Svensson (SV) models, cross-sectional data series are obtained from daily data among August 2, 2011 and December 31, 2021. These data include spot rates obtained from Euribor Overnight Index Swap (OIS) maturing from 1 month up to 11 months, and zero rates derived from applying bootstrapping to Euribor Interest Rates Swap (IRS) with maturities from 1 to 10 years \citep{annaert2013estimating}. Finally, the 30-year Euribor IRS is used to study the out-of-sample extrapolation quality of the different methodologies. 

The data, sourced from the Thompson DataStream, are limited to the end of 2021 due to the replacement of Euribor OIS with Euro Short-Term Rate (€STR) at that date \citep{ECB2020}.

  \begin{table}
  \centering
  \begin{tabular}{p{3cm}p{3cm}p{3cm}p{2cm}p{2cm}}
    \hline
  Maturity& Mean& Std.dev.& Min.& Max. \\
    \hline
  1 month&   -0.1728&0.3226  &-0.5678 &1.2037 \\
  3 months&   -0.1835&0.3119  &-0.5992 &1.2027\\
  6 months&   -0.1932&0.3064  &-0.6347 &1.2302\\
  1 year&   -0.1985&0.3098  &-0.6854 &1.2464\\
  2 years&   0.0664&0.4970 &-0.5712 &1.8354\\
  5 years&   0.3888&0.6331&-0.5533&2.4203\\
  10 years&   0.9616&0.8131 &-0.3323 &3.2260\\
    \hline
  \end{tabular}
  \caption{Descriptive statistics of some spot rates (in percentage).}
  \label{tabla1}
\end{table}

Table \ref{tabla1} summarizes the descriptive statistics for the time series of some rates used to fit the spot rate curve. The table shows that the time series volatility decreases from 0.32\% for the 1-month rates to 0.31\% for the 1-year rates, and then increases to 0.81\% for the 10-years spot rates. The rate average decreases up to 1-year rates, then increases as time to maturity extends. The spot rate curve is generally upward sloping except for the segment up to 1-year maturity. The 10-years rates decreased from 3.23\% in 2011 to 1.48\% in 2013 due to the financial crisis, and further dropped to -0.33\% in 2019 as a result of the COVID crisis. In contrast, the 1-month spot rates remained relatively stable, varying between 1.20\% and almost -0.57\%.

 Note that for every day in our time series, Nelson-Siegel and Nelson-Siegel-Svensson models are estimated using OLS, ridge regression and raise regression. This process results in a total of 2719 models for a whole grid of $\lambda$ values from 0 to 10 with 21 observations for each model. 
    \subsection{The non-essential and essential multicollinearity problem}
 
    As mentioned in the previous sections, it is important to detect whether there is a collinearity problem in both the Nelson-Siegel and Nelson-Siegel-Svensson models. If such a problem exists and both models are estimated using OLS, significant consequences may arise which will be further elaborated upon in the subsequent Section \ref{timeseries}.
       
    \begin{table}
  \centering
  \begin{tabular}{p{3cm}p{3cm}p{3cm}p{2cm}}
    \hline
  & OLS& Ridge& Raise \\
    \hline
   $CN _{NS}$&   48.7774&17.3934  &17.324 \\
    $CN _{SV}$&   67.7693&20.0001  &20.0000 \\
    \hline
  \end{tabular}
  \caption{Mean of the CN with OLS, ridge and raise regression in NS and SV.}
  \label{tabla2}
\end{table}

Observe that the mean of the OLS Condition Number is calculated for both the NS and SV models in Table \ref{tabla2}. Both values exceed the first threshold established as concerning, which is 20, and even surpass the threshold of 30. Therefore, it is evident that both models present a strong multicollinearity problem.

    \begin{table}
  \centering
  \begin{tabular}{p{3cm}p{3cm}p{3cm}p{2cm}}
    \hline
  & CV& VIF \\
    \hline
    $\mathbf{X}_{1 NS}$ & 0.3457 &23.6714 \\
     $\mathbf{X}_{2 NS}$&  0.6589& 23.6714 \\
    $\mathbf{X}_{1 SV}$ &   0.6344&113.2137 \\
     $\mathbf{X}_{2 SV}$&   0.4377&2.6164  \\
     $\mathbf{X}_{3 SV}$ &   0.6725&117.1304  \\
    \hline
  \end{tabular}
  \caption{ Mean of the CV and VIF with OLS in NS and SV.}
  \label{tabla3}
\end{table}

On the other hand, both the mean of the VIF and CV in OLS of each independent variable of the Nelson-Siegel and Nelson-Siegel-Svensson models can be observed in Table \ref{tabla3}. Since all the CVs means of both models are above the threshold established as concerning, which is 0.1002, there would not be a worrying non-essential multicollinearity problem in either model. However, as the VIFs averages of all independent variables (except for the VIF mean of variable $\mathbf{X}_{2 SV}$) are above the threshold of 10, a worrying essential multicollinearity problem would be encountered in both models.

Following the collinearity detection in both models, alternative estimations to OLS are used to treat this problem such as ridge and raise regression. The average value of $k$ selected is 0.0026 and 0.0071 in ridge regression for NS and SV, respectively. In the raise regression the mean value of $k$ selected is 1.5838 and 2.5047 for NS and SV, respectively. To choose which independent variable to raise in the raise regression; the criterion of raising the variable with the highest VIF in SV will be used, namely variable $\mathbf{X}_{3 SV}$. However, as shown in Table \ref{tabla3}, the VIFs means are the same in NS; therefore, the criterion of raising the variable with the lower CV will be used, namely variable $\mathbf{X}_{1 NS}$.

After treating the collinearity issue with ridge and raise regression, it is interesting to verify that this problem has indeed been solved in both models. To do so, the Condition Number mean of the ridge and raise regression are calculated in both models. Note in Table \ref{tabla2} that none of the CNs averages of the ridge and raise regression in the NS and SV models exceed the threshold established as concerning, which is 20. Therefore, the worrying essential collinearity problem has been mitigated.

    \subsection{The time series of estimated parameters}\label{timeseries}

    \begin{figure}[!tb]
      \begin{center}
  \subfigure[$\beta_{0}+\beta_{1}$ (OLS, ridge and raise regression)]{
       \includegraphics[width=0.49\textwidth]{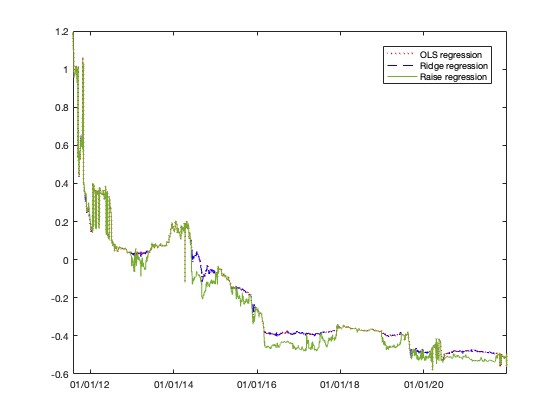}
      \label{beta0beta1ns}}
   \subfigure[$\beta_{0}$ (OLS, ridge and raise regression)]{
       \includegraphics[width=0.49\textwidth]{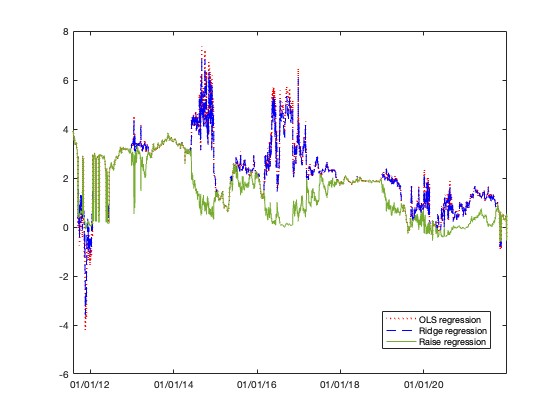}
       \captionsetup{justification=centering}
        \label{beta0ns}}
    \caption{ Time series of long and short term rates estimates using the NS model. }
    \label{parametersns}
    \end{center}
\end{figure}

    \begin{figure}[!tb]
      \begin{center}
  \subfigure[$\beta_{0}+\beta_{1}$ (OLS, ridge and raise regression)]{
       \includegraphics[width=0.49\textwidth]{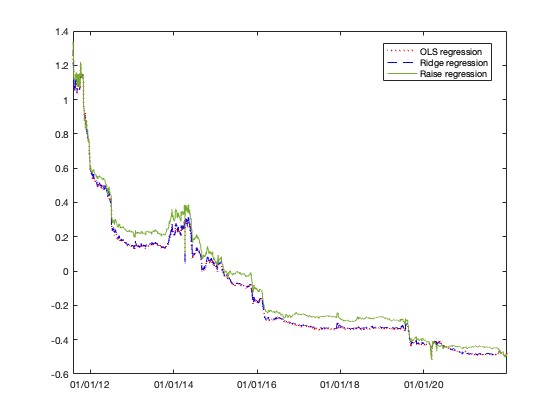}
      \label{beta0beta1sv}}
   \subfigure[$\beta_{0}$ (OLS, ridge and raise regression)]{
       \includegraphics[width=0.49\textwidth]{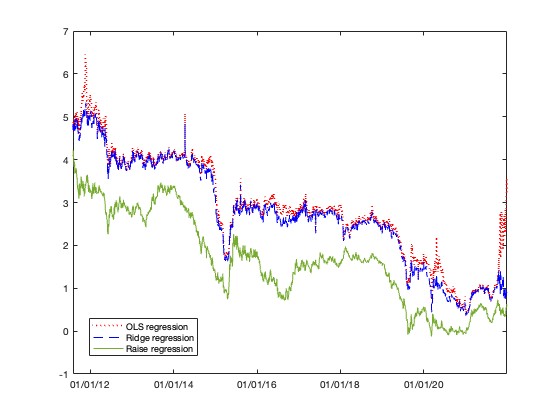}
        \label{beta0sv}}
    \caption{Time series of long and short term rates estimates using the SV model.}
    \label{parameterssv}
    \end{center}
\end{figure}

Figure \ref{parametersns} and  Figure \ref{parameterssv} graphically represent the time series of the long and short term rates estimates with the Nelson-Siegel and Nelson-Siegel-Svensson models, respectively, using OLS, ridge and raise methods. Observe that $\beta_{0}$ coefficient (long-term rate estimate) is notably erratic, there is no clear or predictable trend, using OLS regression both in the NS and SV models at certain points in time. However, the short-term rate estimate ($\beta_{0}+\beta_{1}$) has a more stable behavior using OLS regression both in the Nelson-Siegel and Nelson-Siegel-Svensson models.

On the other hand, the long-term rate estimate is more stable when applying raise regression compared to ridge regression in the NS model (see Figure \ref{beta0ns}), as the latter method still exhibits erratic behavior at certain points in time. Furthermore, raise regression method does not produce extremely smooth time series of the estimated parameters, unlike when the shape parameter is fixed, instead taking a middle position. Therefore, one of the consequences of multicollinearity problem, unstable estimates sensitive to small changes in data, has been resolved with raise regression.

Another consequence of multicollinearity problem is that the signs of the coefficients may be opposite to those expected. For instance, in Figure \ref{beta0ns}, the long-term rates estimates obtained through OLS and ridge regression show highly negative values at the end of 2011. In contrast, the short-term rates estimates using OLS and ridge regression is positive at the same dates. This discrepancy does not make economic sense, as the long-term rate is typically higher than the short-term rate. However, it can be observed that this issue has been resolved with raise regression.
 
    \begin{figure}[!tb]
      \begin{center}
  \subfigure[Nelson-Siegel model]{
       \includegraphics[width=0.49\textwidth]{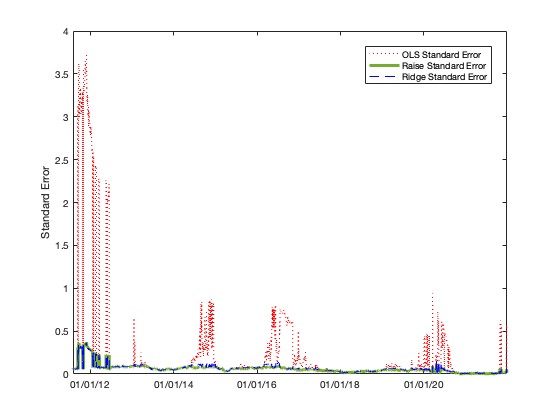}
      \label{bodesns}}
   \subfigure[Nelson-Siegel-Svensson model]{
       \includegraphics[width=0.49\textwidth]{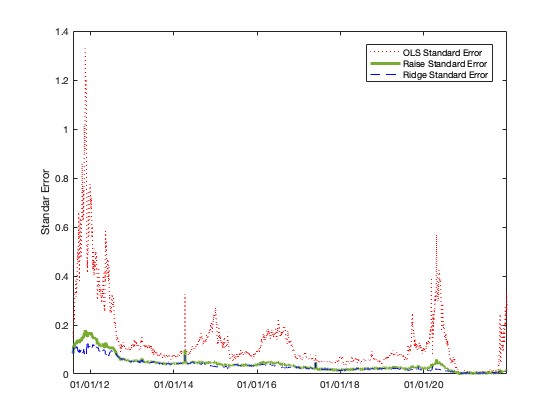}
        \label{bodessv}}
    \caption{ Standard errors of $\beta_{0}$ coefficient with OLS, ridge and raise regression.}
    \label{bodes}
    \end{center}
\end{figure}

OLS estimation not only results in the erratic time series of $\beta_{0}$ coefficient, the precision of the estimates is also very time varying. The Figure \ref{bodes} shows the standard errors of the long-term rates with OLS, ridge and raise regression. Whereas the OLS standard errors are small at times, many periods of turbulence are shown in which the standard errors become high. However, the ridge and raise regression standard errors generally are smaller, obtaining a better accuracy of the estimates and solving another consequence of the multicollinearity problem, inflated coefficient variances.

    \subsection{In-sample performance}

In order to examine the in-sample performance, the Mean Squared Error (MSE) is computed following the expression (\ref{ECM.def}) for OLS, ridge and raise methods in the NS and SV models. 

\begin{table}
  \centering
  \begin{tabular}{p{3cm}p{3cm}p{3cm}p{2cm}}
    \hline
  & OLS& Ridge& Raise \\
    \hline
    $MSE _{NS}$&   0.0379&0.0380  &0.0379 \\
     $MSE _{SV}$&   0.0132&0.0136  &0.0132 \\
    \hline
  \end{tabular}
  \caption{In-sample Mean Squared Error with OLS, ridge and raise regression using the NS and SV models.}
  \label{tabla4}
\end{table}

 Table \ref{tabla4} shows that the fit of both estimated models to the real sample values using the three methods (OLS, raise regression and ridge regression) is very similar, with very little difference between MSEs obtained. However, the Nelson-Siegel and Nelson-Siegel-Svensson models exhibit a worse fit using ridge regression compared to raise regression, with the difference being greater in the SV model than in the NS model by 0.04 and 0.01 percentage points, respectively. Note that, as mentioned in Section \ref{raiseregression}, MSE obtained using raise regression in both models is the same as that obtained using OLS. Although mitigating the multicollinearity problem usually increases bias, this bias does not change in the case of raise regression.
 
    \subsection{Out-of-sample performance}
  In order to examine the out-sample performance, the Mean Absolute Errors (MAE) between the estimated 30-year swap rate and 30-year Euribor IRS is computed as follows:
  \begin{equation}
    \label{MAE.def}
    MAE= \frac{1}{m} \sum^{m}_{j=1} |r_{j, 30}-\hat{r}_{j, 30}|,
\end{equation}
where $r_{j, 30}$ are the real values and $\hat{r}_{j, 30}$ are the predicted values. The 30-year spot rates estimated through the NS and SV models are transformed into swap rates by following the methodology shown in \cite{annaert2013estimating}.

Then,  Table \ref{tabla5} can be observed to investigate the ability of these three approaches to extrapolate the long ends of the term structure with Nelson-Siegel and Nelson-Siegel-Svensson models.
     
     \begin{table}
  \centering
  \begin{tabular}{p{3cm}p{3cm}p{3cm}p{2cm}}
    \hline
  & OLS& Ridge& Raise \\
    \hline
  $MAE _{NS}$&   0.1192&0.1053  &0.0380 \\
   $MAE _{NS}$&   0.3436&0.2575  &0.0244 \\
    \hline
  \end{tabular}
  \caption{Out-of-sample Mean Absolute Errors with OLS, ridge and raise regression using the NS and SV models.}
  \label{tabla5}
\end{table}

      \begin{figure}[!tb]
      \begin{center}
  \subfigure[Nelson-Siegel model]{
       \includegraphics[width=0.49\textwidth]{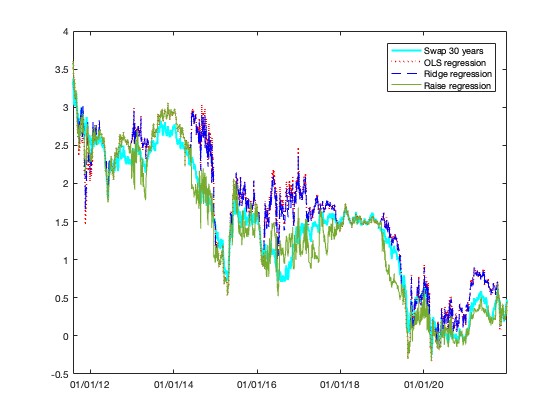}
      \label{swapns}}
   \subfigure[Nelson-Siegel-Svensson model]{
       \includegraphics[width=0.49\textwidth]{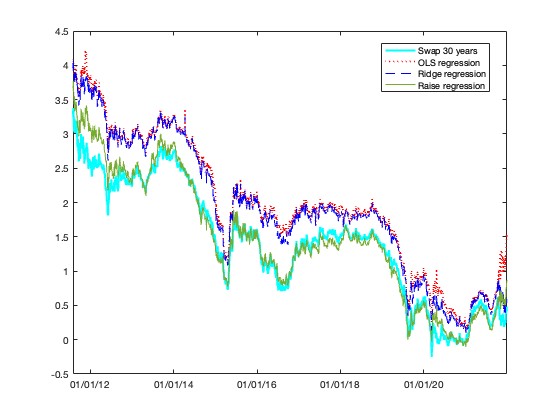}
        \label{swapsv}}
    \caption{ 30-year Euribor IRS and estimated swap rate using OLS, ridge and raise regression.}
    \label{swap}
    \end{center}
\end{figure}

In this case, both the Table \ref{tabla2} and the Figure \ref{swap} show that the raise regression has the lowest MAE when extrapolating the long-term rate in both the NS and SV models. The MAE is reduced by 8 and 32 percentage points in the Nelson-Siegel and Nelson-Siegel-Svensson models, respectively, compared with OLS MAE. Furthermore, the MAE obtained from ridge regression is also lower compared to OLS. However, this difference is much smaller, with 1 and 9 percentage points in the Nelson-Siegel and Nelson-Siegel-Svensson models, respectively,

Finally, it is tested whether the MAEs obtained in Table 5 are significantly different from each other (with a 95\% confidence interval). To do so, the MAEs are compared using the following expressions and the t-test with Newey-West correction on the standard errors to remove the posible serial correlation from the residuals:

\begin{equation}
    \label{statistically.def1}
   \hat{\alpha}_{1}  = MAE_{OLS}- MAE_{raise},
\end{equation}

\begin{equation}
    \label{statistically.def2}
   \hat{\alpha}_{2}= MAE_{ridge}- MAE_{raise},
\end{equation}

\begin{equation}
    \label{statistically.def3}
   \hat{\alpha}_{3} = MAE_{OLS}- MAE_{ridge},
\end{equation}

 \begin{table}
  \centering
  \begin{tabular}{p{3cm}p{3cm}p{3cm}p{2cm}}
    \hline
  &$   t-test_{1} $& $  t-test_{2} $& $   t-test_{3} $ \\
    \hline
$t-test_{NS}$& 8.1447  &7.7667  &8.5484 \\
   $t-test_{SV}$&   45.4712&  39.3709&15.4100 \\
    \hline
  \end{tabular}
  \caption{t-test with $\hat{\boldsymbol{\alpha}}$ and Newey-West correction on the standard errors.}
  \label{tabla6}
\end{table}

The results in Table \ref{tabla6} show significant differences between all calculated MAEs for both the NS and SV models at a 5\% significance level. Furthemore, all the estimated coefficients, $\hat{\boldsymbol{\alpha}}$, are significantly positive, indicating that the raise method is superior to both OLS and ridge methods. Finally, ridge regression is shown to be better than OLS.

\section{Conclusions}  \label{conclusiones}

Several studies have shown that there could be a collinearity problem in the Nelson-Siegel and Nelson-Siegel-Svensson models because the independent variables are related to each other. When collinearity problem appears, the coefficient estimates in this models using  OLS  can become very unstable, with inflated variances and signs opposite to those expected. However, these models are used to estimate the term structure by European Central Banks and other agents to make important decisions, such as monetary policy issues. For this reason, it is essential to mitigate the collinearity problem.

For this purpose some traditional methods are used as: nonlinear optimization, which is very sensitive to the initial value used; fixing the shape parameter over the whole time series, which can result in an extremely smooth time series of the parameters estimates; and ridge regression, which reduces the coefficient of determination and requires standardized data, potentially losing an important economic parameter, $\beta_{0}$. Therefore, this paper proposes raise regression, which maintains the original characteristics of the model (in terms of goodness of fit and global significance) and  does not require data standardization, while effectively mitigating collinearity.

To illustrate the raise method and compare it with OLS and ridge regression, the Euribor spot rate curve is estimated 2719 times for the period 2011-2021 using both the Nelson-Siegel and the Nelson-Siegel-Svensson models. The OLS estimation results in erratic time series for the coefficient $\beta_{0}$ in both the Nelson-Siegel and Nelson-Seigel-Svensson models. Additionally, some values of the long-term rate estimate are highly negative in the Nelson-Siegel model, which makes no economic sense. Finally, the precision of the estimates is highly variable over time, exhibiting numerous periods of turbulence where the standard errors reach significantly high values. 

The ridge estimation also exhibits an unstable time series for the coefficient $\beta_{0}$ in the Nelson-Siegel model, and some long-term rate estimates are highly negative in this same model. However, the estimated time series for the coefficient $\beta_{0}$ is more stable when applying raise regression, taking a middle position; the issue of highly negative long-term rate estimates does not exist and the standard errors are lower during these periods of turbulence, resulting in greater accuracy of the estimates.

On the other hand, the in-sample comparison shows that the fitting of the estimated Nelson-Siegel and Nelson-Siegel-Svensson models to the real sample values using the three methods (OLS, raise regression and ridge regression) is very similar. However, raise regression presents a better fit than ridge regression in both models and the same MSE as OLS.  Although mitigating collinearity increases bias, in the case of raise regression this does not happen.

Finally, the out-of-sample comparison shows that the raise regression has a better extrapolation power the long ends of the term structure both Nelson-Siegel and Nelson-Siegel-Svensson models. The MAEs are lowered with 8 and 32 percentage points in the Nelson-Siegel and Nelson-Siegel-Svensson models, respectively, compared with OLS MAEs. The ridge regression also have a lower MAEs than OLS but this difference is significantly smaller with 1 and 9 percentage points in the Nelson-Siegel and Nelson-Siegel-Svensson models, respectively.

\section*{Acknowledgements}

This work has been supported by project A-SEJ-496-UGR20 of the Andalusian Government's Counseling of
Economic Transformation, Industry, Knowledge and Universities (Spain).
\bibliographystyle{spbasic}      
\bibliography{bib}

\begin{thebibliography}{39}
\providecommand{\natexlab}[1]{#1}
\providecommand{\url}[1]{{#1}}
\providecommand{\urlprefix}{URL }
\expandafter\ifx\csname urlstyle\endcsname\relax
  \providecommand{\doi}[1]{DOI~\discretionary{}{}{}#1}\else
  \providecommand{\doi}{DOI~\discretionary{}{}{}\begingroup
  \urlstyle{rm}\Url}\fi
\providecommand{\eprint}[2][]{\url{#2}}

\bibitem[{Annaert et~al.(2013)Annaert, Claes, De~Ceuster, and
  Zhang}]{annaert2013estimating}
Annaert J, Claes AG, De~Ceuster MJ, Zhang H (2013) Estimating the spot rate
  curve using the nelson--siegel model: A ridge regression approach.
  International Review of Economics \& Finance 27:482--496

\bibitem[{Annaert et~al.(2015)Annaert, Claes, De~Ceuster, and
  Zhang}]{annaert2015estimating}
Annaert J, Claes AG, De~Ceuster MJ, Zhang H (2015) Estimating the long rate and
  its volatility. Economics Letters 129:100--102

\bibitem[{Barrett et~al.(1995)Barrett, Gosnell, and Heuson}]{barrett1995yield}
Barrett WB, Gosnell TF, Heuson AJ (1995) Yield curve shifts and the selection
  of immunization strategies. The Journal of Fixed Income 5(2):53--64

\bibitem[{Belsley(1991)}]{belsley1991guide}
Belsley DA (1991) A guide to using the collinearity diagnostics. Computer
  Science in Economics and Management 4(1):33--50

\bibitem[{Besley et~al.(1980)Besley, Kuh, and Welsch}]{besley1980regression}
Besley D, Kuh E, Welsch R (1980) Regression Diagnostics: Identifying
  Influential Data and Sources of Collinearity,. New York: John Wiley and Sons

\bibitem[{Cairns and Pritchard(2001)}]{cairns2001stability}
Cairns AJ, Pritchard DJ (2001) Stability of models for the term structure of
  interest rates with application to german market data. British Actuarial
  Journal 7(3):467--507

\bibitem[{Cera et~al.(2020)Cera, Philippe, and Vladimir}]{ECB2020}
Cera K, Philippe M, Vladimir T (2020) Some way to go in the transition to the
  €str. Financial Stability Review

\bibitem[{Choudhry(2010)}]{choudhry2010fixed}
Choudhry M (2010) Fixed-income securities and derivatives handbook: Analysis
  and valuation. John Wiley \& Sons

\bibitem[{Cox et~al.(1985)Cox, Ingersoll~Jr, and Ross}]{cox1985intertemporal}
Cox JC, Ingersoll~Jr JE, Ross SA (1985) An intertemporal general equilibrium
  model of asset prices. Econometrica: Journal of the Econometric Society pp
  363--384

\bibitem[{Diebold and Li(2006)}]{diebold2006forecasting}
Diebold FX, Li C (2006) Forecasting the term structure of government bond
  yields. Journal of econometrics 130(2):337--364

\bibitem[{Fabozzi et~al.(2005)Fabozzi, Martellini, and
  Priaulet}]{fabozzi2005predictability}
Fabozzi FJ, Martellini L, Priaulet P (2005) Predictability in the shape of the
  term structure of interest rates. The Journal of Fixed Income 15(1):40--53

\bibitem[{Farrar and Glauber(1967)}]{Farrar1967}
Farrar DE, Glauber RR (1967) Multicollinearity in regression analysis: the
  problem revisited. The Review of Economic and Statistics pp 92--107

\bibitem[{Friedman(1977)}]{friedman1977time}
Friedman M (1977) Time perspective in demand for money. The Scandinavian
  Journal of Economics 79(4):397--416

\bibitem[{Garc{\'\i}a et~al.(2011)Garc{\'\i}a, Garc{\'\i}a, and
  Soto}]{garcia2011raise}
Garc{\'\i}a C, Garc{\'\i}a J, Soto J (2011) The raise method. an alternative
  procedure to estimate the parameters in presence of collinearity. Quality \&
  Quantity 45(2):403--423

\bibitem[{Garc{\'\i}a et~al.(2016)Garc{\'\i}a, Salmer{\'o}n, Garc{\'\i}a, and
  L{\'o}pez~Mart{\'\i}n}]{garcia2016standardization}
Garc{\'\i}a J, Salmer{\'o}n R, Garc{\'\i}a C, L{\'o}pez~Mart{\'\i}n MdM (2016)
  Standardization of variables and collinearity diagnostic in ridge regression.
  International Statistical Review 84(2):245--266

\bibitem[{Gilli et~al.(2010)Gilli, Gro{\ss}e, and
  Schumann}]{gilli2010calibrating}
Gilli M, Gro{\ss}e S, Schumann E (2010) Calibrating the nelson-siegel-svensson
  model. Available at SSRN 1676747

\bibitem[{Gunst and Mason(1977)}]{Gunst1977b}
Gunst RF, Mason RL (1977) Advantages of examining multicollinearities in
  regression analysis. Biometrics pp 249--260

\bibitem[{Hoerl and Kennard(1970{\natexlab{a}})}]{hoerl1970ridgeb}
Hoerl AE, Kennard RW (1970{\natexlab{a}}) Ridge regression: applications to
  nonorthogonal problems. Technometrics 12(1):69--82

\bibitem[{Hoerl and Kennard(1970{\natexlab{b}})}]{hoerl1970ridgea}
Hoerl AE, Kennard RW (1970{\natexlab{b}}) Ridge regression: Biased estimation
  for nonorthogonal problems. Technometrics 12(1):55--67

\bibitem[{James and Webber(2000)}]{james2000interest}
James J, Webber N (2000) Interest rate modelling. Wiley-Blackwell Publishing
  Ltd.

\bibitem[{Marquardt(1970)}]{marquaridt1970generalized}
Marquardt DW (1970) Generalized inverses, ridge regression, biased linear
  estimation, and nonlinear estimation. Technometrics 12(3):591--612

\bibitem[{McCulloch(1971)}]{mcculloch1971measuring}
McCulloch JH (1971) Measuring the term structure of interest rates. the Journal
  of Business 44(1):19--31

\bibitem[{Nelson and Siegel(1987)}]{nelson1987parsimonious}
Nelson CR, Siegel AF (1987) Parsimonious modeling of yield curves. Journal of
  business pp 473--489

\bibitem[{Rodr\'iguez et~al.(2019)Rodr\'iguez, Salmer\'on, and
  Garc\'ia}]{Rodriguez2019}
Rodr\'iguez A, Salmer\'on R, Garc\'ia C (2019) The coefficient of determination
  in the ridge regression. Communications in Statistics - Simulation and
  Computation \urlprefix\url{10.1080/03610918.2019.1649421}

\bibitem[{Rodr{\'\i}guez et~al.(in press)Rodr{\'\i}guez, Garc{\'\i}a, and
  Salmer{\'o}n}]{nelsoncapitulo}
Rodr{\'\i}guez A, Garc{\'\i}a CB, Salmer{\'o}n R (in press) Nelson-siegel model
  and multicollinearity. Springer Nature

\bibitem[{Rold{\'a}n et~al.(2020)Rold{\'a}n, Garc{\'\i}a, and
  Salmer{\'o}n}]{de2020analysis}
Rold{\'a}n AL, Garc{\'\i}a C, Salmer{\'o}n R (2020) Analysis of the condition
  number in the raise regression. Communications in Statistics-Theory and
  Methods pp 1--16

\bibitem[{Salmer{\'o}n et~al.(2017{\natexlab{a}})Salmer{\'o}n, Garcia, Garcia,
  and Lopez}]{salmeron2017raise}
Salmer{\'o}n R, Garcia C, Garcia J, Lopez MM (2017{\natexlab{a}}) The raise
  estimator estimation, inference, and properties. Communications in
  Statistics-Theory and Methods 46(13):6446--6462

\bibitem[{Salmer{\'o}n et~al.(2017{\natexlab{b}})Salmer{\'o}n, Garc{\'\i}a,
  Garc{\'\i}a, and Mart{\'\i}n}]{salmeron2017note}
Salmer{\'o}n R, Garc{\'\i}a J, Garc{\'\i}a C, Mart{\'\i}n ML
  (2017{\natexlab{b}}) A note about the corrected vif. Statistical Papers
  58(3):929--945

\bibitem[{Salmer{\'o}n et~al.(2018)Salmer{\'o}n, Garc{\'\i}a, and
  Garc{\'\i}a}]{salmeron2018variance}
Salmer{\'o}n R, Garc{\'\i}a C, Garc{\'\i}a J (2018) Variance inflation factor
  and condition number in multiple linear regression. Journal of Statistical
  Computation and Simulation 88(12):2365--2384

\bibitem[{Salmer\'on et~al.(2018)Salmer\'on, Garc\'ia, and
  Garc\'ia}]{Salmeron2018Garcia}
Salmer\'on R, Garc\'ia C, Garc\'ia J (2018) Variance inflation factor and
  condition number in multiple linear regression. Journal of Statistical
  Computation and Simulation 88(12):2365--2384

\bibitem[{Salmer{\'o}n et~al.(2018)Salmer{\'o}n, Garc{\'\i}a, Garc{\'\i}a, and
  L{\'o}pez}]{salmeron2018transformation}
Salmer{\'o}n R, Garc{\'\i}a J, Garc{\'\i}a C, L{\'o}pez MdM (2018)
  Transformation of variables and the condition number in ridge estimation.
  Computational Statistics 33:1497--1524

\bibitem[{Salmer{\'o}n-G{\'o}mez et~al.(2019)Salmer{\'o}n-G{\'o}mez,
  Rodr{\'\i}guez-S{\'a}nchez, and
  Garc{\'\i}a-Garc{\'\i}a}]{salmeron2019noesencial}
Salmer{\'o}n-G{\'o}mez R, Rodr{\'\i}guez-S{\'a}nchez A, Garc{\'\i}a-Garc{\'\i}a
  C (2019) Diagnosis and quantification of the non-essential collinearity.
  Computational Statistics pp 1--20

\bibitem[{Salmer{\'o}n-G{\'o}mez et~al.(2024)Salmer{\'o}n-G{\'o}mez,
  Garc{\'\i}a-Garc{\'\i}a, and Garc{\'\i}a-P{\'e}rez}]{gomez2021raise}
Salmer{\'o}n-G{\'o}mez R, Garc{\'\i}a-Garc{\'\i}a CB, Garc{\'\i}a-P{\'e}rez J
  (2024) The raise regression: Justification, properties and application.
  International Statistical Review

\bibitem[{Shea(1984)}]{shea1984pitfalls}
Shea GS (1984) Pitfalls in smoothing interest rate term structure data:
  Equilibrium models and spline approximations. Journal of Financial and
  Quantitative Analysis 19(3):253--269

\bibitem[{Svensson(1994)}]{svensson1994estimating}
Svensson LE (1994) Estimating and interpreting forward interest rates: Sweden
  1992-1994. Tech. rep., National bureau of economic research

\bibitem[{Valle et~al.(2018)Valle, Serrano, and
  Marco}]{leon2018multicollinearity}
Valle AL, Serrano AR, Marco LS (2018) On multicollinearity and the value of the
  shape parameter in the term structure nelson-siegel model. Aestimatio: The
  IEB International Journal of Finance (16):8--29

\bibitem[{Vasicek(1977)}]{vasicek1977equilibrium}
Vasicek O (1977) An equilibrium characterization of the term structure. Journal
  of financial economics 5(2):177--188

\bibitem[{Virmani(2012)}]{virmani2012estimability}
Virmani V (2012) On estimability of parsimonious term structure models: An
  experiment with the nelson--siegel specification. Applied Economics Letters
  19(17):1703--1706

\bibitem[{Willan and Watts(1978)}]{willan1978meaningful}
Willan AR, Watts DG (1978) Meaningful multicollinearity measures. Technometrics
  20(4):407--412

\end{thebibliography}

\end{document}